\documentclass[11pt]{iopart}
\usepackage{amstext}
\usepackage{amssymb}
\usepackage{graphicx}
\usepackage{color}


\begin{document}


\title[Quantum correlations and spatial localization in 1D ultracold bosonic mixtures \dots]
{Quantum correlations and spatial localization in one-dimensional ultracold bosonic mixtures}

\author{M. A. Garcia-March$^1$, B. Juli\'a-D\'iaz$^{1,2}$,
G.~E. Astrakharchik$^3$, Th. Busch$^4$, J. Boronat$^3$, and A. Polls$^1$}
\address{$^1$Departament d'Estructura i Constituents
de la Mat\`eria, Univ. de Barcelona, 08028 Barcelona, Spain}
\address{$^2$  ICFO-Institut de Ci\`encies Fot\`oniques,
Parc Mediterrani de la Tecnologia, 08860 Barcelona, Spain}
\address{$^3$Departament de F\'\i sica i Enginyeria Nuclear,
Campus Nord B4, Universitat Polit\`ecnica de Catalunya, E-08034 Barcelona, Spain}
\address{$^4$Quantum Systems Unit, OIST Graduate University, Okinawa, Japan}

\begin{abstract}
We present the complete phase diagram for one-dimensional binary mixtures of bosonic ultracold atomic gases in a harmonic trap. We obtain exact results with direct numerical diagonalization for small number of atoms, which permits us to quantify quantum many-body correlations. The quantum Monte Carlo method is used to calculate energies and density profiles for larger system sizes. We study the system properties for a wide range of interaction parameters. For the extreme values of these parameters, different  correlation limits can be identified, where the  correlations are either weak or strong.
We investigate in detail how the correlation evolve between the limits.
For balanced mixtures in the number of atoms in each species, the transition between the different limits involves sophisticated changes in the one- and two-body correlations. Particularly, we  quantify the entanglement between the two components by means of the von Neumann entropy. We show that the limits equally exist when the number of atoms is increased, for balanced mixtures. Also, the changes in the correlations along the transitions among these limits are qualitatively similar. We also show that, for imbalanced mixtures, the same limits with similar transitions exist. Finally, for strongly imbalanced systems, only two limits survive, i.e., a miscible limit and a phase-separated one, resembling those expected with a mean-field approach.

\end{abstract}

\maketitle

\section{Introduction}

The fascinating physics of interpenetrating superfluids has recently become a topic of large interest due to the experimental realisation of multi-component, atomic Bose-Einstein condensates \cite{Myatt:97,StamperKurn:98,Hall:98,Catani:08,McCarron:11}.  In the weakly interacting regime, these mixtures are well described by coupled mean-field Gross-Pitaevskii equations (GPEs), and within this framework processes that lead to phase separation are well described~\cite{Ho:96,Law:97,Esry:97,Timmermans:98,Pu:98,Goldstein:97,Busch:97,Ao:98,Trippenbach:00}

While mean-field theories allow to study weakly correlated systems, it is also important and interesting to examine  quantum mixtures in strongly correlated regimes. In these regimes, analytic solutions can often only be obtained in limiting cases. Rather appealing  results occur in strongly correlated regimes when the dimensionality is reduced. For quasi one-dimensional (1D) gas mixtures one finds that Luttinger liquid theory predicts many interesting effects, which include de-mixing for repulsive interactions or spin-charge separation analogous to that found in 1D electronic quantum systems~\cite{Cazalilla:03,Alon:06,Mishra:07,Kleine:08}.  Other relevant effects include the presence of polarized ground states, which allow to view the relative spatial oscillations as spin waves~\cite{Ho:98,Eisenberg:02,Fuchs:05,Guan:07} and which have been experimentally observed~\cite{Lewandowski:02,McGuirk:02,McGuirk:03}.

Very strong correlations for single component bosons are realized in the Tonks-Girardeau (TG) gas~\cite{Girardeau:60, Girardeau:01,Gangardt:03}, which was recently observed experimentally~\cite{Paredes:04,Kinoshita:04}. Bosonic mixtures in the strongly interacting limit have features common with the TG gas, and their ground-state wavefunction can be obtained
analytically in certain interaction limits~\cite{Girardeau:07,Deuretzbacher:08,Zinner:13}.
Experimental advances on Feshbach and confined induced resonances in recent years have made it possible
to control both, the intra-species interactions
and the inter-species interactions, over a wide range of parameters~\cite{Olshanii1998,Papp:08,Thalhammer:08}. In the strongly interacting
limit a number of relevant phenomena have been described including
phase separation~\cite{Cazalilla:03, Alon:06,Mishra:07,Garcia-march:12},
composite fermionization~\cite{Zollner:08a,Hao:09,Hao:09b}, a sharp crossover
between both limits ~\cite{Garcia-March:13}, and quantum magnetism~\cite{Deuretzbacher:13}.

In this work we focus on mixtures where the number of atoms is small. The recent
successful experimental trapping of ensembles of few
atoms~\cite{He:10,Serwane:11,Wenz:13,Bourgain:13} has inspired
an intense theoretical effort in few-atom
systems~\cite{Busch:98,Idziaszek:06,Kestner:07,Pflanzer:2009,Pflanzer:2010,Liu:10,
Blume:12,Gharashi:12,Harshman:12,DAmico:13,Harshman:13,Sowinski:13,Volosniev:13,Wilson:13}.
For mixtures of few atoms, direct diagonalization
methods~\cite{Deuretzbacher:07,Hao:09b,Garcia-march:12}, can be used
together with other numerical methods efficient for larger numbers of atoms, like multiconfigurational Hartree-Fock
methods (MCTDH)~\cite{Alon:07}, density functional theory (DFT)\cite{Hao:09},
or quantum diffusion  Monte Carlo (DMC)~\cite{Boronat:94}. In the present work, we use
direct numerical diagonalization to study the
ground-state properties of a mixture of ultracold bosons confined in a 1D trap over a wide range of correlations regimes, determined by the scattering properties between the atoms. These are supplemented by DMC calculations to confirm trends for systems with larger particle numbers. While the extreme cases in which all correlations are either weak or strong are well known, here we calculate and discuss the full phase diagram and especially the transitions between the different regimes.

We study the ground-state wavefunction, and pay particular attention to the one- and
two-body correlations in the extreme limits, and  across the transitions between them.
The quantum correlations between both components are characterized
by means of the von Neumann entropy.  This allows us to show that close
to the crossover between the {\it composite fermionization} and {\it phase separation},
the ground state exhibits strong correlations between the two bosonic components.

Our manuscript is organized as follows. In Sec.~\ref{Sec:model} we introduce
the model Hamiltonian and a general analytical ansatz for the ground-state wavefunction.  Focusing first on balanced mixtures, we discuss in Sec.~\ref{Sec:BalancedMixtures} the ground-state properties in terms of the densities, the coherence, the energies, the one- and two-body correlations, and the von Neumann entropy.
In Sec.~\ref{Sec:weaklylarger}, we then present results on how the ground-state properties change when one component is larger than the other and finally summarize all our results in Sec.~\ref{Sec:conc}.

\begin{figure}[ht]
  \begin{center}
  \includegraphics[width=0.9\linewidth]{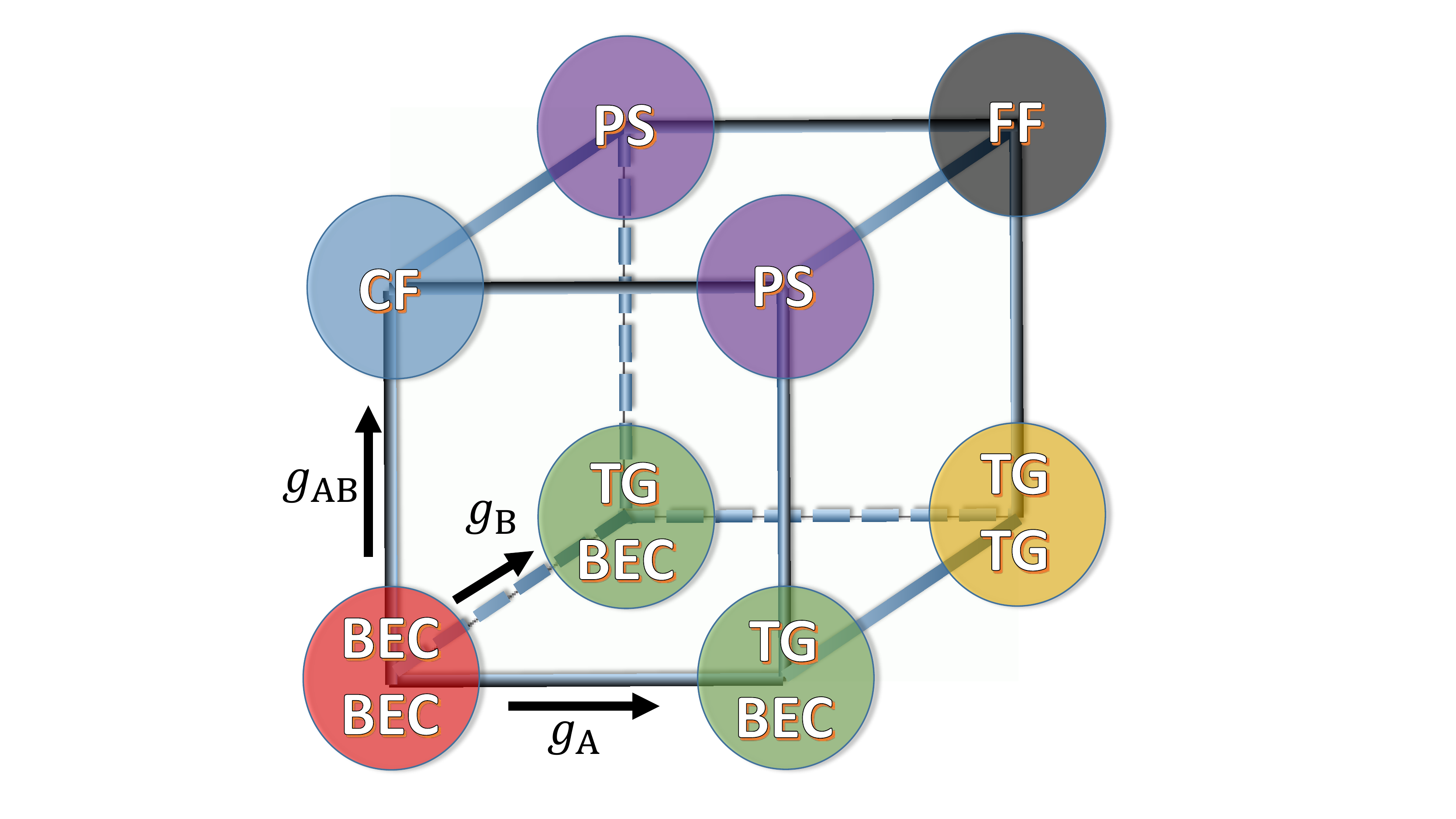}
  \end{center}
  \caption{(Color online) Schematic of all regimes in the few atom limit.
  \label{Fig0}}
\end{figure}

 \section{Model Hamiltonian}
 \label{Sec:model}

Let us consider a mixture of two bosonic components, A and B, with a small,
fixed number of atoms in each component, $N_{\mathrm{A}}$ and $N_{\mathrm{B}}$.  We assume that
the two components are two different hyperfine states of the same atomic species
of mass $m$, and that they are trapped in the same, one-dimensional parabolic
potential $V(x)=\frac{1}{2}m\omega^{2}x^{2}$. At low temperatures, all
scattering processes between the atoms are assumed to  be described
by contact interactions
$v_{\mathrm{int}}^\mathrm{A}=g_\mathrm{A}\delta(x_j-x_{j'}) $,
$v_{\mathrm{int}}^\mathrm{B}=g_\mathrm{B}\delta(y_j-y_{j'}) $, and
$ v_{\mathrm{int}}^\mathrm{AB}=g_\mathrm{AB}\delta(x_i-y_{j})$, where the positions of
atoms of species A(B) are given by the coordinates $x_j(y_j)$. The 1D
intra- and inter-species coupling constants $g_\mathrm{A(B)}$ and $g_\mathrm{AB}$
are assumed to be tunable independently by means of confinement induced
resonances~\cite{Olshanii1998}. We will restrict our study to repulsive interactions.  The many-body Hamiltonian is
$\hat \mathcal H= \hat \mathcal H_\mathrm{A}+\hat \mathcal H_\mathrm{B} +\hat \mathcal H_\text{int}$,
with
\begin{eqnarray}
& \hat \mathcal H_\mathrm{A}=\sum_{j=1}^{N_\mathrm{A}}\left[-\frac{\hbar^2}{2m}\frac{\partial^2}{\partial x_j^2}+V(x_j) \right]+\sum_{j<j'}^{N_\mathrm{A}}v_{\mathrm{int}}^\mathrm{A}(x_j,x_{j'}),\nonumber\\
 &\hat \mathcal H_{B}=\sum_{j=1}^{N_\mathrm{B}}\left[-\frac{\hbar^2}{2m}\frac{\partial^2}{\partial y_j^2}+V(y_j) \right]+\sum_{j<j'}^{N_\mathrm{B}}v_{\mathrm{int}}^\mathrm{B}(y_j,y_{j'}),\nonumber\\
 &\hat \mathcal H_\text{int} =\sum_{j=1}^{N_\mathrm{A}} \sum_{j'=1}^{N_\mathrm{B}} v_{\mathrm{int}}^{\mathrm{AB}}(x_j-y_{j'}).
\label{eq:GHamiltonian}
\end{eqnarray}

There are three coupling constants $g_\mathrm{A}, g_\mathrm{B}, g_{\mathrm{AB}}$ each of them ranging from $g=0$ for ideal Bose gas interaction to $g\to\infty$ for strong Tonks-Girardeau interaction. This defines eight limits schematically shown in Fig.~\ref{Fig0}. The composite fermionization limit is reached when $g_{\mathrm{AB}}\to \infty$ with the other coupling constants vanishing~\cite{Zollner:08a,Hao:09,Hao:09b}. We termed TG-BEC gas a system with one of the  intra-species coupling constants large, while other coupling constants vanish~\cite{Garcia-march:12}. If one of the  intra-species coupling constants together with the inter-species coupling constant are large, the phase separation limit is reached~\cite{Cazalilla:03, Alon:06,Mishra:07,Garcia-march:12}. Finally, if all coupling constants tend to infinity, the wavefunction is known exactly and can be mapped to the one of an ideal Fermi gas~\cite{Girardeau:07}. We call this limit full fermionization.  In the following we will calculate and discuss 
the complete phase diagram, which 
includes the transitions between these limits.  
To restrict the large number of free parameters, we note the transition between TG and phase separation limit is symmetric when switching the values of $g_\mathrm{A}$ and $g_\mathrm{B}$ and we can therefore circumscribe the discussion to the situation where $g_\mathrm{B}$ is small and change $g_\mathrm{A}$.
In the following, we will use harmonic oscillator units and scale all lengths in units of oscillator length $a_0=\sqrt{\hbar/(m\omega)}$ and all energies in units of level spacing $\hbar\omega$.

To solve the Hamiltonian (\ref{eq:GHamiltonian}) we use two different numerical approaches: direct diagonalization~\cite{Garcia-march:12} and  DMC~\cite{Boronat:94}. The former allows us to calculate the full  density matrix of the system and therefore gives us access to all single and multi-particle correlations. However, since it is limited to small particle numbers, the latter will be used to check for trends when the number of particles becomes larger.  While DMC is well described in the literature, let us briefly explain our approach to direct diagonalization. For this we expand the second quantised field operators into
eigenfunctions, $\phi_{n}(x)$, of the single-particle (SP) Hamiltonian  for
the harmonic oscillator
\begin{equation}
   \hat{\psi}_\mathrm{A}(x)=\sum_{n=1}^{n_\mathrm{A}}\hat{a}_n\phi_n(x),\qquad\text{and}\qquad
   \hat{\psi}_\mathrm{B}(x)=\sum_{n=1}^{n_\mathrm{B}}\hat{b}_n\phi_n(x) \,,
\end{equation}
where the creation and annihilation operators, $\hat{a}_{k}^{\dagger}$ and $\hat{a}_{k}$,
satisfy the bosonic commutation relations $[\hat{a}_{k},\hat{a}_{l}^{\dagger}]=\delta_{kl}$,
$[\hat{a}_{k},\hat{a}_{l}]=[\hat{a}_{k}^{\dagger},\hat{a}_{l}^{\dagger}]=0$, and similarly for
$\hat{b}_{k}^{\dagger}$ and $\hat{b}_{k}$, while all commutators between operators belonging
to different species vanish. Here, $n_\mathrm{A(B)}$ is the number of modes used in the expansion. The Hamiltonian can then be written as~\cite{Garcia-March:13}
\begin{eqnarray}
\label{eq:Hamiltonian}
 & \hat H_\mathrm{A} \!=\!\sum_{k}\hat{a}^\dagger_k\hat{a}_k \hbar\omega\left(\frac{1}{2}+k\right)+\frac{1}{2}
        \sum_{klmn}\hat{a}_k^\dagger \hat{a}_l^\dagger \hat{a}_m \hat{a}_n V_{klmn}^\mathrm{A}\\
 &\hat H_\mathrm{B} \!=\!\sum_{k}\hat{b}^\dagger_k\hat{b}_k \hbar\omega\left(\frac{1}{2}+k\right)+\frac{1}{2}
        \sum_{klmn} \hat{b}_k^\dagger \hat{b}_l^\dagger \hat{b}_m \hat{b}_n V_{klmn}^\mathrm{B}\\
 &\hat H_{\mathrm{int}}\!=\!\sum_{klmn} \hat{a}_k^\dagger \hat{b}_l^\dagger \hat{b}_m \hat{a}_n V_{klmn}^\mathrm{AB}\,,
\end{eqnarray}
where
\begin{eqnarray}
 \label{eq:parameters}
V_{klmn}^\mathrm{A(B)}=&g_\mathrm{A(B)}\int dx\;\phi_k^*(x)\phi_l^*(x)\phi_m(x)\phi_n(x),\\
V_{klmn}^\mathrm{AB}=&g_\mathrm{AB}\int dx\;\phi_k^*(x)\phi_l^*(x)\phi_m(x)\phi_n(x) \,.
\end{eqnarray}
The ground state can be expressed in terms of Fock vectors
$\Psi_{0}=\sum_{i=1}^{\Omega}c_{i}\Phi_{i}$ with
\begin{equation}
\label{eq:FockBasis}
 \Phi_i=D_i^{\mathrm{A}}D_i^{\mathrm{B}}\left(\hat{a}_{1}^\dagger\right)^{N_{1,i}^\mathrm{A}}\!\!\!\!\dots
         \left(\hat{a}_{n_\mathrm{A}}^\dagger\right)^{N_{n_\mathrm{A},i}^\mathrm{A}}
         \left(\hat{b}_{1}^\dagger\right)^{N_{1,i}^\mathrm{B}}\!\!\!\!\dots
         \left(\hat{b}_{n_{B}}^\dagger\right)^{N_{n_\mathrm{B},i}^\mathrm{B}}\Phi_0,
\end{equation}
where $D_i^{A(B)}=(N_{1,i}^\mathrm{A(B)}!\dots N_{n_\mathrm{A(B)},i}^\mathrm{A(B)}!)^{-\frac{1}{2}}$ and $\Phi_{0}$ is the vacuum. The occupation numbers of the $n_\mathrm{A}$ ($n_{B}$) modes for each component are given by  $N_{1,i}^\mathrm{A},\dots,N_{n_\mathrm{A},i}^\mathrm{A}$ ($N_{1,i}^\mathrm{B},\dots,N_{n_{B},i}^\mathrm{B}$). The dimension of the Hilbert space  is
$\Omega=\Omega_\mathrm{A}\Omega_{B}$ with
$\Omega_\mathrm{A(B)}=(N_\mathrm{A(B)}+n_\mathrm{A(B)}-1)!/N_\mathrm{A(B)}!(n_\mathrm{A(B)}-1)!$.
Note that $ \Omega$ increases exponentially
with the number of particles and modes, which is the reason why the numerical solution using this
approach is restricted to small numbers of atoms.

A good ansatz for the unnormalized ground-state wavefunction of the mixture when $g_{\mathrm{B}}=0$ and outside of the phase-separated regime  can be constructed using the solution for non-interacting atoms in the harmonic trap, $\Phi(X)= \exp [-\sum x_i^2 / 2  ]$, $X=\{x_i\}$ and $Y=\{y_i\}$, as~\cite{Garcia-March:13}
\begin{eqnarray}
\label{wavefunc0}
\Psi(X,Y)&=\Phi(X) \, \Phi(Y) \prod_{j<k}^{N_{\mathrm{A}}}|x_k-x_j-a_{\mathrm{A}}|
\quad \prod_{k}^{N_{\mathrm{A}}}\prod_{j}^{N_{\mathrm{B}}}|x_k-y_j-a_{\mathrm{AB}}|.
\end{eqnarray}
Here the 1D $s$-wave scattering length $a_ \mathrm{A}$ for the intra-species interactions and $a_ \mathrm{AB}$ for the inter-species interactions
are related to the 1D coupling constants as $g_{\mathrm{A}}=-2\hbar^2/(m a_{\mathrm{A}})$ and $g_{\mathrm{AB}}=-2\hbar^2/(m a_{\mathrm{AB}})$ and we assume that both coupling constants are non-negative corresponding to repulsive interactions. For practical purposes, we find that the coupling constant  $g=20$  is close enough to the infinite limit, and therefore we use this value in the direct diagonalization method in describing the large coupling constant limits.

\section{Balanced Mixtures}
 \label{Sec:BalancedMixtures}

In the following we will first concentrate on systems in which both components have the same particle number. Unless otherwise stated, we will use $N_\text{A}=N_\text{B}=2$.

\subsection{Densities}

\begin{figure}[tb]
\includegraphics[scale=0.43]{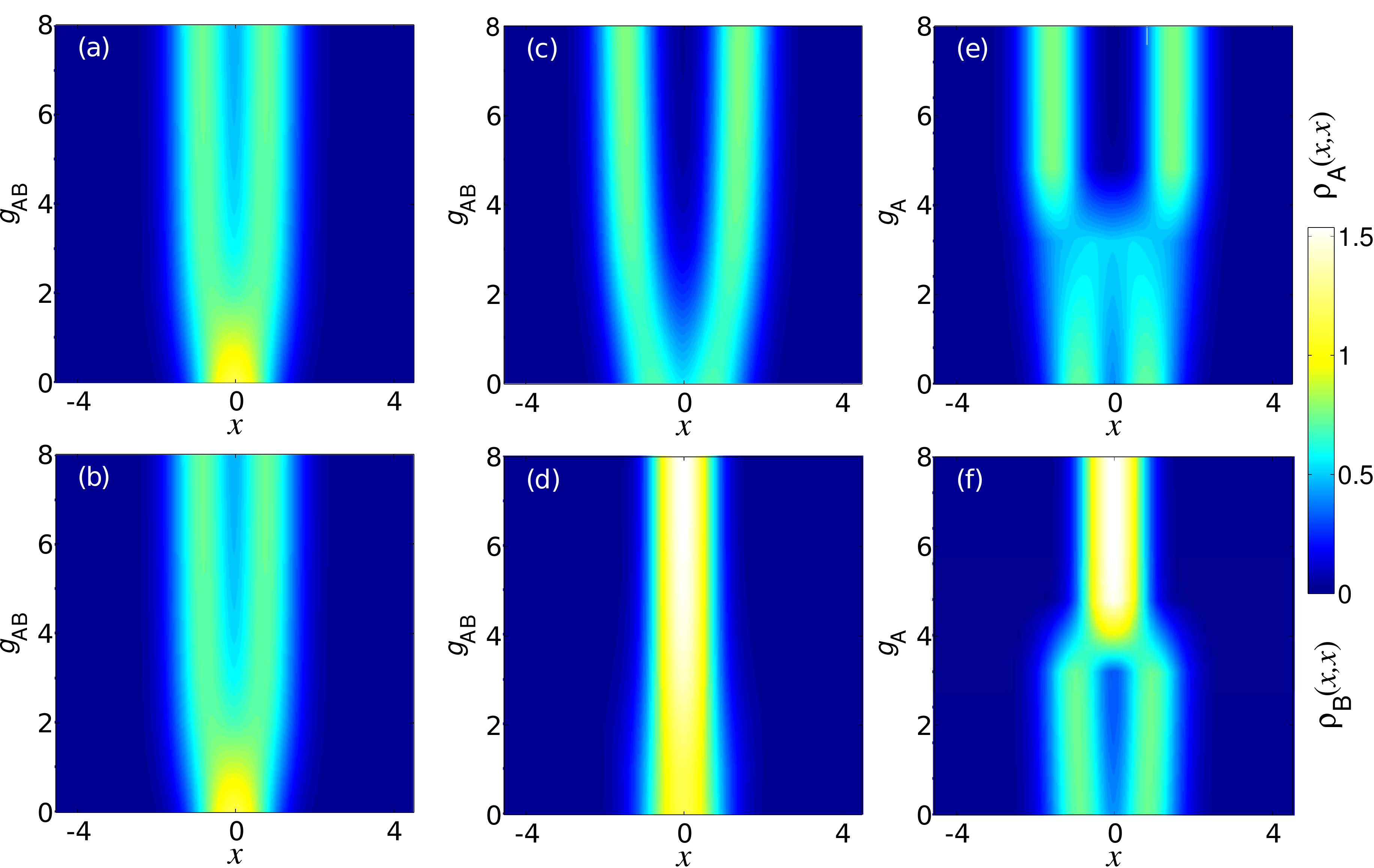}
  \caption{(Color online) Upper (lower) row shows the density of the A (B) species, for $N_\mathrm{A}=N_\mathrm{B}=2$.  Panels (a-d) show the evolution for increasing $g_\text{AB}$, starting from the BEC-BEC limit [panels (a) and (b), $g_\text{A}=g_\text{B}=0$] or the TG-BEC limit [panels (c) and (d), $g_\text{A}=20$, $g_\text{B}=0$].  Panels (e) and (f) display the transition between the composite fermionization and the phase-separated limits [$g_\text{B}=0$, $g_\mathrm{AB}=20$].
  \label{Fig:Densities}}
\end{figure}

The main feature of the density evolution in this system is the occurrence of phase separation for increasing inter-species interactions. However this process takes two, fundamentally different forms: in the composite fermionization limit atoms of different species avoid each other even though the species' densities still occupy the same space, whereas in the phase separation limit the overlap of the respective densities goes to zero.

The density along the transition from the BEC-BEC limit (all couplings small) to the composite fermionization limit ($g_\mathrm{AB}$ large) is shown in Figs.~\ref{Fig:Densities}(a-b). There are crucial differences in the evolution of the density along the transition from the TG-BEC to the phase separation limit (Figs.~\ref{Fig:Densities}(c-d)). One immediately notices that the transition into the composite fermionization state happens at a finite value of $g_\text{AB}\sim2$, whereas the transition to the phase-separated regime happens already for very small values of $g_\text{AB}$. Also the final state reached in the composite fermionization or the phase separation limit are very different.

This difference in the final states can be understood by looking at the one-body density matrix (OBDM) given by
\begin{eqnarray}
\label{OBDM}
 \rho_1^\mathrm{A}(x,x')&=
N_\mathrm{A}\int dx_2\,\cdots\,dx_{N_\mathrm{A}}\,dy_1\,\cdots\,dy_{N_\mathrm{B}}|\Psi|^2 \\
&=\sum_{k}f_{k}(x)f_{k}(x')
\lambda_{\mathrm{A}}^k \label{OBDMnat}
\end{eqnarray}

with a similar expression for $\rho_1^\mathrm{B}(x,x')$. The decomposition in terms of natural orbitals $f_{k}(x)$ of the OBDM and their corresponding occupations $\lambda_{\mathrm{A}}^k $ is given in Eq.~(\ref{OBDMnat}).  The densities shown in Fig.~\ref{Fig:Densities} are the diagonals of these matrices, calculated with direct diagonalization. As discussed in Ref.~\cite{Garcia-March:13}, the OBDM of both components in the composite fermionization limit are identical and show two peaks. Contrary, in the phase separation limit the OBDM of B shows a single peak located at the center of the trap, while the OBDM for A shows two peaks at the edges. The largest used value of the coupling constant $g=8$ is big enough, so that the density profiles shown in Figs.~\ref{Fig:Densities} are practically the same as for the infinite coupling constant.

Finally, the transition from the composite fermionization to the phase-separated regime is shown in Figs.~\ref{Fig:Densities}(e) and (f). One can see that the spatial separation of the clouds happens for a finite value of $g_\text{A}$. At the transition between  both limits, the OBDM of both species show a complicated structure, which we discuss in detail in subsec.~\ref{sec:corrematrices}.

\subsection{Coherence and Entanglement}

\begin{figure}[tb]
 \begin{center}
 \includegraphics[scale=1.15]{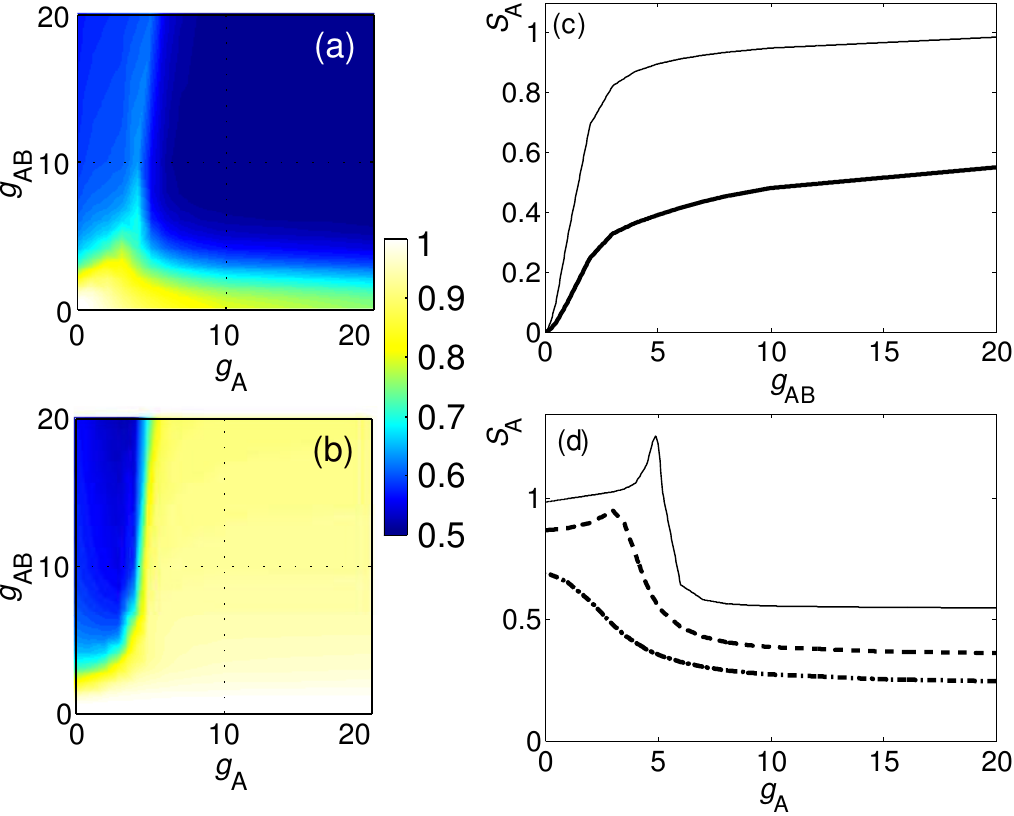}
 \end{center}
\caption{(Color online)  Largest occupation numbers of the natural orbitals for (a) the A species, $\lambda_0^\mathrm{A}$, and (b) the B species, $\lambda_0^\mathrm{B}$.  (c)  von Neumann entropy for $N_{\mathrm{A}}=N_{\mathrm{B}}=2$ as a function of $g_{\mathrm{AB}}$ for $g_{\mathrm{A}}=0$ (thick line) and  $g_{\mathrm{A}}\rightarrow\infty$ (thin line).  (d)  von Neumann entropy as a function of $g_{\mathrm{A}}$ for the cases $g_\text{AB}=2,4,20$ (dash-dotted, dashed, and solid line, respectively) for $N_{\mathrm{A}}=N_{\mathrm{B}}=2$.
  \label{Fig:Coherence}}
\end{figure}

Since increasing the coupling constant will drive the system from the weakly to the strongly correlated regime, the coherence is a good quantity for identifying different regions in the phase diagram. It can be characterised by the largest eigenvalue of the OBDMs~(\ref{OBDMnat}), $\lambda^{\mathrm{A(B)}}_0$, which provides the  largest occupation of a natural orbital. In our numerical calculations with direct diagonalization we normalize the OBDM to 1 instead of the number of atoms. In Figs.~\ref{Fig:Coherence}(a) and (b) we show the largest occupation numbers for the A and the B species, respectively, over the whole range of interactions. Note that the sum of all eigenvalues of each component sum up to 1, in accordance with the chosen normalization.

One can see from Fig.~\ref{Fig:Coherence}(a)  that the coherence in the A species decreases monotonically along the transition from the  BEC-BEC  ($\lambda^{\mathrm{A}}_0=1$) to the  TG-BEC ($\lambda^{\mathrm{A}}_0\sim 0.7$) limit, as well as to the composite fermionization limit ($\lambda^{\mathrm{A}}_0\sim0.55$). However, the transition for increasing $g_\mathrm{A}$ at a finite $g_\mathrm{AB}$ shows that a maximum of coherence is reached for finite values of $g_\mathrm{A}\sim 5$, which corresponds roughly to the value where the cloud de-mixing happens (see Figs.~\ref{Fig:Densities}(e) and (f)). This maximum in coherence within species A is very surprising, as usually the presence of interactions is thought of as detrimental to coherence. Here, however the presence of interactions within the A component to a certain degree ``counterbalance'' the interactions between the species and therefore allows to re-establish a higher degree of coherence again. Note that after the de-mixing transition the coherence 
within species A
goes down again, which is a clear
indication that the enhancement is somehow mediated using the overlap with species B.

As expected, species B shows a large degree of coherence in all limits, except the composite fermionization one (see Fig.~\ref{Fig:Coherence}(b)) . However, the re-establishment of coherence along the transition from composite fermionization to the phase-separated limit happens over a definite and narrow region, which corresponds to the area in which the coherence in species A shows a maximum.

One might, at this point wonder how the transition to phase separation manifests itself during the transition from the TG-BEC to the phase-separated limit, as no obvious signature is visible in the coherence phase diagram. The answer is that phase separation happens already for small values of $g_\mathrm{AB}$, which can be seen in Figs.~\ref{Fig:Densities}(c). 

It is important to observe that there are no phase transitions in the whole phase diagram. The ground-state energy is always a continuous and smooth function of the parameters, so that the transition between the different regimes is of crossover type. 

Closely related to the coherence in the sample is the entanglement between the two components. This can be quantified by calculating the von Neumann entropy, $S_A=-\mathrm{Tr}\left(\rho_A\ln\rho_A\right)$, which is a function of the reduced density matrix for a single component %
\begin{equation}
\label{trace}
\rho_A=\mathrm{Tr}_B\rho=\sum_i\langle\Phi_i^{\mathrm{B}}|\Psi_{0}\rangle\langle \Psi_{0}|\Phi_i^{\mathrm{B}}\rangle.
\end{equation}
Here $\rho=|\Psi_{0}\rangle\langle \Psi_{0}|$ is the density matrix, $\Psi_{0}$
is the system ground state, and  
\begin{equation}
\Phi_i^{\mathrm{B}} = D_i^{\mathrm{B}}\left(\hat{b}_1^\dagger\right)^{N_{1,i}^\mathrm{B}}\!\!\!\!\dots
         \left(\hat{b}_{n_{B}}^\dagger\right)^{N_{n_\mathrm{B},i}^\mathrm{B}}\Phi_0,
\end{equation}
is the Fock vector for species B only.  This matrix is obtained by means of  direct
diagonalization. 
%
%
In Fig.~\ref{Fig:Coherence}(c) we show the
von Neumann entropy $S_A$  along the transition between
BEC-BEC and composite fermionization.  $S_A$ can be seen to approach a constant
value as $g_\mathrm{AB}$ is increased,  corresponding  to the large inter-species correlations present in the
composite fermionization.
The same plot also shows $S_A$ along the transition between TG-BEC and phase-separated limit. 
The two species are less correlated throughout this transition, but still $S_A$ saturates to a constant value in the phase-separated  limit. In
Fig.~\ref{Fig:Coherence}(d) we plot $S_A$ for different values of $g_\mathrm{AB}$ when
$g_\mathrm{A}$ is tuned from zero to a large value. When $g_\mathrm{AB}=20$,
this corresponds to the transition between composite fermionization and a
phase-separated  gas. We observe a peak which coincides with the
crossover between both limits. This peak disappears as $g_\mathrm{AB}$ is
reduced, as observed in the curves for $g_\mathrm{AB}=4,2$ in  Fig.~\ref{Fig:Coherence}(d) (for $g_\mathrm{AB}=0$, $S_A$ is zero for every value of $g_\mathrm{A}$).

 \subsection{Interaction Energies}

\begin{figure}[tb]
 \begin{center}
\includegraphics[scale=1.15]{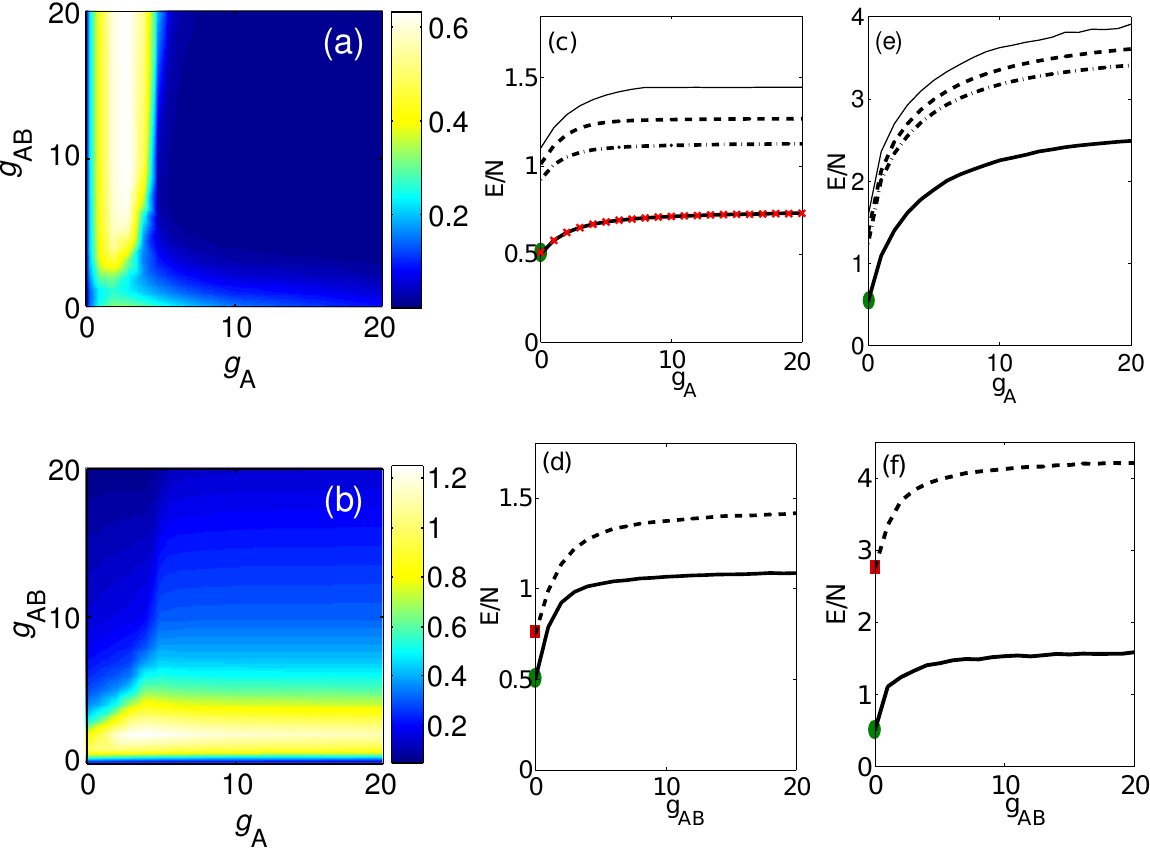}
 \end{center}
\caption{(Color online)
Panel (a) shows the average interaction energy of species A, $\langle U_{\mathrm{A}}\rangle$ and panel (b) the average interaction energy  between species A and B, $\langle U_{\mathrm{AB}}\rangle$. Here $N_\mathrm{A}=N_\mathrm{B}=2$ and $g_\mathrm{B}=0$. Panel (c) reports the  energy per atom as a function of $g_{\mathrm{A}}$ for $g_{\mathrm{AB}}=0,2,4,20$ (thick solid, dash-dotted, dashed, and thin solid lines, respectively) for $N_{\mathrm{A}}=N_{\mathrm{B}}=2$.  The red crosses overlapping with the black thick line represent  the analytical result~\cite{Busch:98}. (d) Energy per atom as a function of $g_{\mathrm{AB}}$ for $g_{\mathrm{A}}=0$ (solid line) and  $g_{\mathrm{A}}\rightarrow\infty$ (dashed line), for $N_{\mathrm{A}}=N_{\mathrm{B}}=2$. Panels (e) and (f) represent the energy per atom for $N_{\mathrm{A}}=N_{\mathrm{B}}=10$, with the same layout than figures (c) and (d), respectively. In  panels (c) to (f) the green circles indicate the energy in the BEC-BEC limit. In panels (d) and (f) the red 
squares indicate the
energy in the  TG-BEC limit.
  \label{Fig:Energies}}
\end{figure}

An interesting question is how the interaction energy changes across the transitions between the different limits. The  average interaction energy in species A is
\begin{equation}
  \langle U_\mathrm{A}\rangle =\left\langle \frac{1}{2}
\sum_{klmn}\hat{a}_k^\dagger \hat{a}_l^\dagger \hat{a}_m \hat{a}_n V_{klmn}^\mathrm{A}\right\rangle.
\end{equation}
We  display this energy in Fig.~\ref{Fig:Energies}(a). For zero $g_\mathrm{A}$ there are no interactions between A atoms and $\langle U_\mathrm{A} \rangle$ is equal to zero. By increasing $g_\mathrm{A}$ the energy $\langle U_\mathrm{A}\rangle$ first grows as correlations are being introduced. For larger repulsion, particles avoid each other which leads to very strong correlation and the interaction energy drops down to zero. Starting from the BEC-BEC region, this is a long drawn process, however for a finite value of $g_\text{AB}$ this happens over a very well defined domain of the parameter $g_\text{A}$,  located at  small values of $g_\text{A}$. Note that for $g_\text{A}=0$ and in the presence of interaction with species B  the particles in species A are much more localised than for $g_\text{AB}=0$. Therefore, small increases in the interaction strength  $g_\text{A}$ leads to strong increases in the interaction energy $ \langle U_{\mathrm A}\rangle$. This is also consistent with the maximum found in the 
correlation strength within component A.

The interaction energy goes to zero in the TG-BEC limit, which is the behaviour expected for a single component  gas~\cite{Alon:05,Deuretzbacher:07,Ernst:11,Brouzos:12}, as the increased energy is now stored in the single particle harmonic oscillator energies. During the
whole process the total energy is increased from 
\begin{equation}
E_\mathrm{BECBEC}=\frac{1}{2}(N_\mathrm{A}+N_\mathrm{B})
\label{E:BECBEC}
\end{equation}
to 
\begin{equation}
E_\mathrm{TGBEC}=\frac{1}{2}(N_\mathrm{B}+N_\mathrm{A}^2).
\label{E:TGBEC}
\end{equation}
The energy for $N_\mathrm{A}=N_\mathrm{B}=2$ is shown in Fig.~\ref{Fig:Energies}(c). 
The energy obtained by the direct diagonalization and DMC methods coincides. 
For no interactions between different species, $g_\text{AB} = 0$, the energy can be expressed as $E = \hbar\omega + E_2(g_\mathrm{A})$, where $E_2(g_\mathrm{A})$ is the energy of two trapped particles interacting with the coupling constant $g_\mathrm{A}$~\cite{Busch:98}. 
In order to prove that the described limits exist in larger systems, we calculate the energy for $N_\mathrm{A}=N_\mathrm{B}=10$ particles with DMC method.  

The energy per particle in the BEC-BEC limit (\ref{E:BECBEC}) does not depend on the number of particles, $E_\mathrm{BECBEC}/N = 1/2$. We show it in Figs.~\ref{Fig:Energies}(c,e) with green circles for $N_A=N_B=2$ and 10.

In Figs.~\ref{Fig:Energies}(d,f) we depict the energy per particle as a function of $g_\mathrm{AB}$, starting from the BEC-BEC (solid line) and the TG-BEC (dahsed line) limits. Here the green circles  (red squares) indicate the energy per atom in the BEC-BEC (TG-BEC) limit. The energy in the TG-BEC limit given by Eq. (\ref{E:TGBEC}) is $E_\mathrm{TGBEC}/ N=3/4$ for $N_A=N_B=2$ and $E_\mathrm{TGBEC}/ N=11/4$ for $N_A=N_B=10$.  In the transition from the BEC-BEC limit to the composite fermionization one, the energy saturates to certain value, for which we do not have an analytical prediction. 
As well, a monotonic behavior is observed in the transition from the TG-BEC to the phase separation limit (Figs.~\ref{Fig:Energies} (d) and (f)).

The average interaction energy between both species, given by
\begin{equation}
  \langle U_\mathrm{AB}\rangle =
\left\langle\sum_{klmn} \hat{a}_k^\dagger \hat{b}_l^\dagger \hat{b}_m \hat{a}_n V_{klmn}^{\mathrm{AB}}\right\rangle,
\end{equation}
is important to quantify the transition to the composite fermionization or the phase-separated regime.  
The interaction energy rapidly increases from zero to a maximum at $g_\mathrm{AB}\approx 2$ (see Fig.~\ref{Fig:Energies}(b)) and decreases again towards zero for $g_\mathrm{AB}\rightarrow\infty$.  
For $g_\mathrm{A}= 0$ this corresponds to building up strong correlations between the particles of different species in the composite fermionization limit, whereas in the limit of large $g_\mathrm{A}$ this reflects the transition to {\it macroscopic} phase separation of the two components.

\subsection{Correlation Matrices}
\label{sec:corrematrices}

\begin{figure*}
\begin{tabular}{ccccc}
\hspace{-0.65cm}
\includegraphics[scale=0.66]{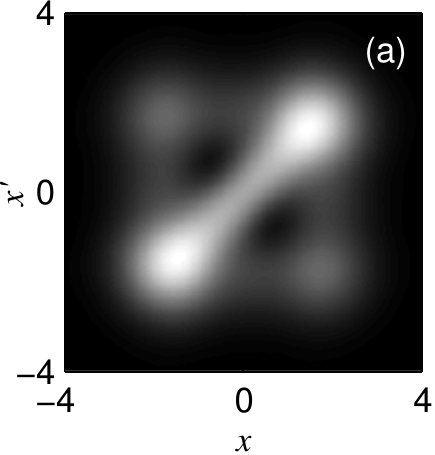} &
\hspace{-.4cm}\includegraphics[scale=.65]{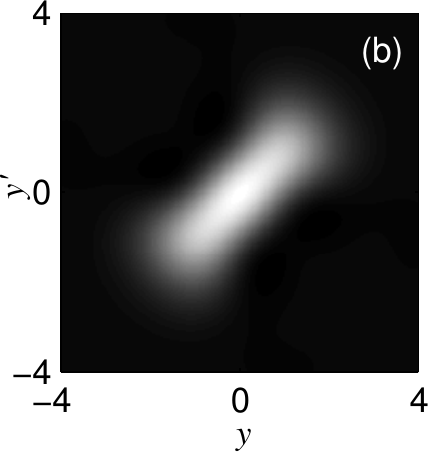} &
\hspace{+.3cm}\includegraphics[scale=.65]{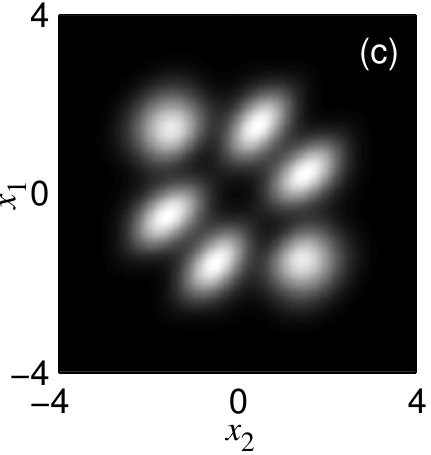} &
\hspace{-.5cm}\includegraphics[scale=.65]{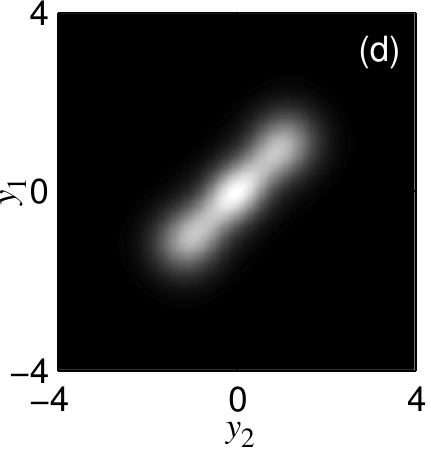} &
\hspace{+.3cm}\includegraphics[scale=.65]{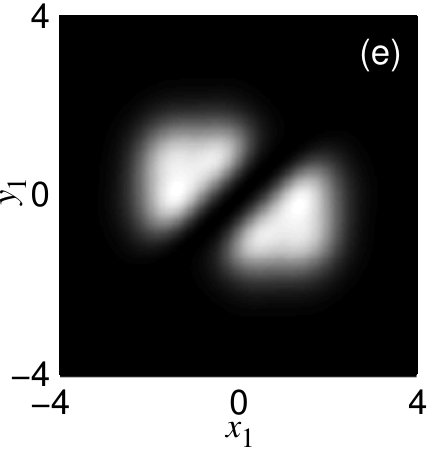}\\
\hspace{-0.5cm}\includegraphics[scale=.65]{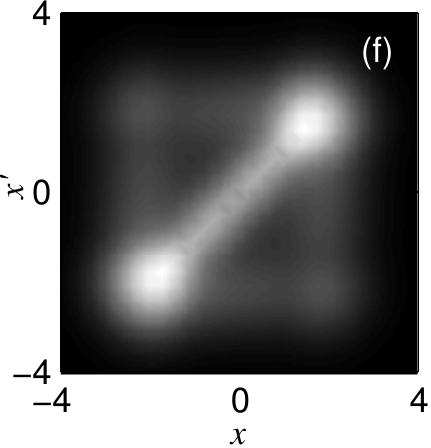} &
\hspace{-.5cm}\includegraphics[scale=.65]{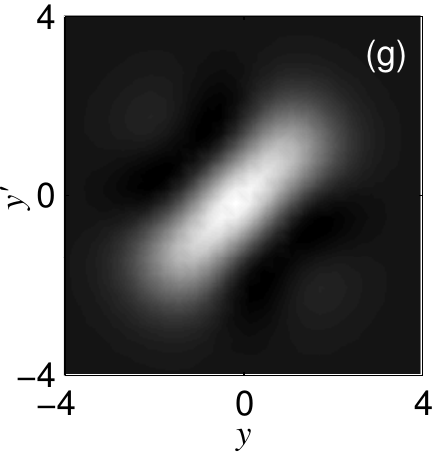} &
\hspace{+.3cm}\includegraphics[scale=.65]{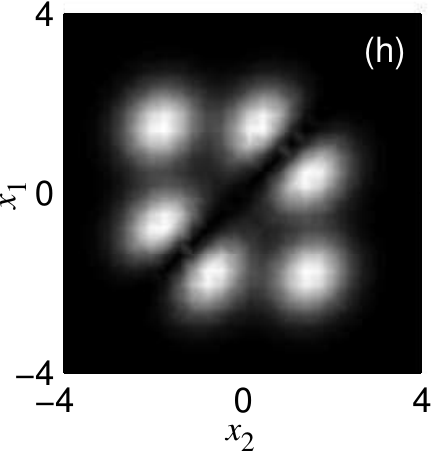} &
\hspace{-.5cm}\includegraphics[scale=.65]{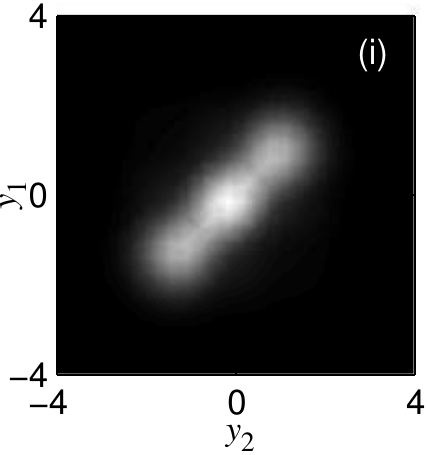} &
\hspace{+.3cm}\includegraphics[scale=.65]{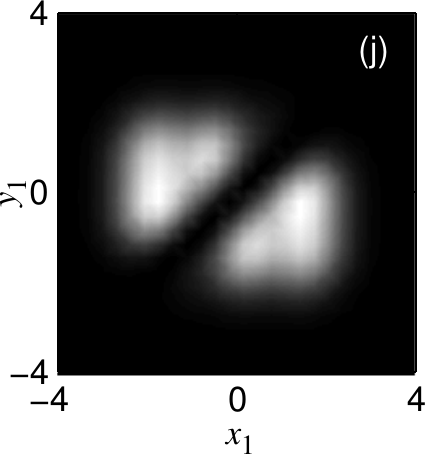}\\
\hspace{-0.5cm}\includegraphics[scale=.65]{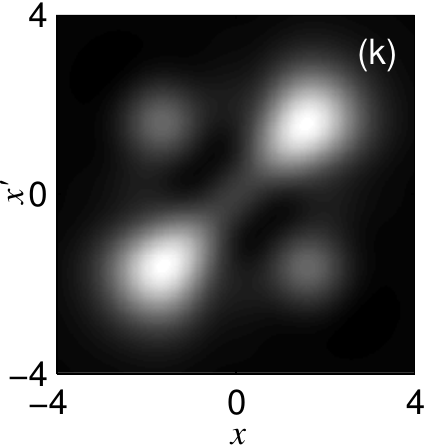} &
\hspace{-.5cm}\includegraphics[scale=.65]{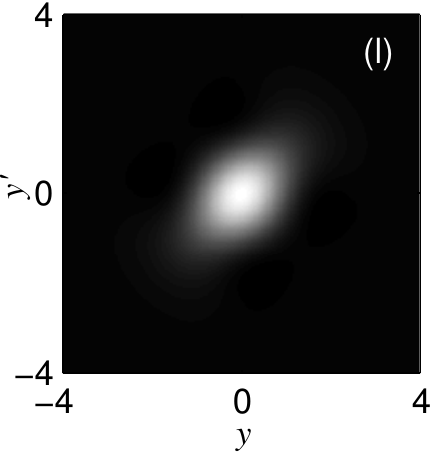} &
\hspace{+.3cm}\includegraphics[scale=.65]{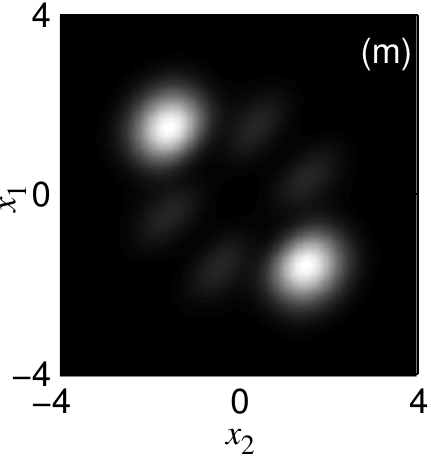} &
\hspace{-.5cm}\includegraphics[scale=.65]{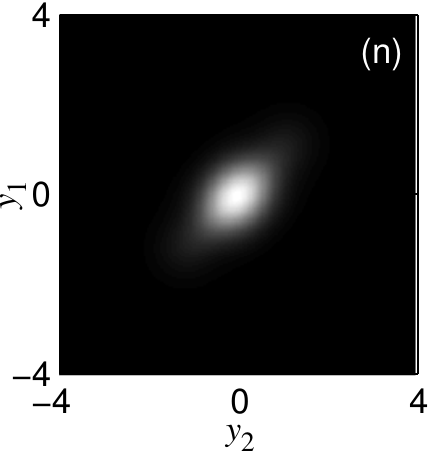} &
\hspace{+.3cm}\includegraphics[scale=.65]{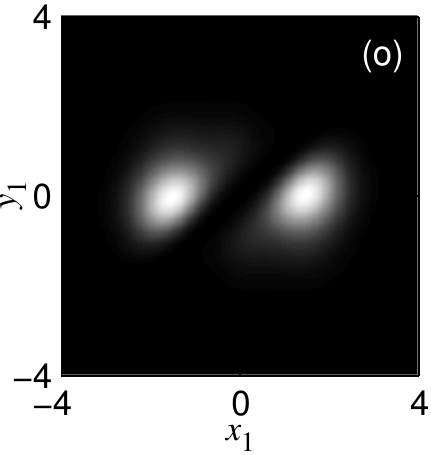}
\end{tabular}
\caption{(Color online)
The first and second columns show the OBDMs, the third and fourth columns the TBDFs, each time for species A and B, respectively, and the last column shows the CTBDF.  $g_\mathrm{AB}$ is large in all panels.
The first and last row displays the numerical result obtained for a value of $g_\mathrm{A}$  just before and after the crossover, respectively, and the middle row  shows the results obtained from calculating
the OBDMs and TBDF directly using the ansatz given in Eq.~(\ref{wavefunc0}) with $a_\mathrm{A}=a_\mathrm{AB}=0$. Good agreement is clearly visible with the numerical results before the crossover. \label{fig4}}
\end{figure*}

Since in the presence of strong interactions the system has non-trivial many-body correlations, it is interesting to look not only at single-particle densities, but also at pair-wise correlation functions. The single-particle densities are quantified by the OBDM, Eq.~(\ref{OBDM}). For the particles of the same species, the two-particle correlations are quantified by the two-body distribution function (TBDF)
\begin{eqnarray}
& \rho_2^\mathrm{A}(x_1,x_2)=N_{\mathrm{A}} (N_{\mathrm{A}}-1)
\int\! dx_3\cdots dx_{N_{\mathrm{A}}} dy_1\cdots dy_{N_{\mathrm{B}}}\,|\Psi|^2,
\end{eqnarray}
with an analogous expression for B. If the two atoms stem from different species, their pair-wise correlations are captured in the cross  two-body distribution function (CTBDF) given by
\begin{eqnarray}
\rho_2^\mathrm{AB}(x_1,y_1)\!=
\!N_{\mathrm{A}}\,N_{\mathrm{B}}\!\int\! dx_2\cdots dx_{N_{\mathrm{A}}}dy_2\cdots dy_{N_{\mathrm{B}}}\,|\Psi|^2.
\end{eqnarray}
Both functions are proportional to the joint probability for finding two atoms at
two given positions. 

It was shown in Ref.~\cite{Zollner:08a, Garcia-March:13} that the correlation functions are very useful for a description of the composite fermionization and the phase separation limits. In the following we will carefully look at the transition between these two limits.  
The phase separation occurs for $g_\mathrm{A}$ and $g_\mathrm{AB}$ large and implies a density distribution with atoms of species B are localized at the center of the trap, while the atoms of species A gather at the edges of the density of B. As discussed above, the de-mixing point can also be identified in the coherence, the interaction energies and the entanglement.


In Fig.~\ref{fig4} we show the OBDMs, TBDFs and CTBDFs just before ($g_\mathrm{A}=5$) and just after ($g_\mathrm{A}=7$)
the crossover. The upper row and lower row show numerical results while the middle row represents the analytical results obtained from  ansatz~(\ref{wavefunc0}) with $a_\mathrm{A}=a_\mathrm{AB}=0$. 
One can see that just before the crossover the densities of both species, i.e.~the diagonals of the OBDMs, significantly overlap (panel (a) and (b)), 
whereas the overlapping is greatly reduced after the crossover (panels (k) and (l)). 
The TBDFs and CTBDFs before and after the crossover (panels (c) to (e) and (m) to (o), respectively) demonstrate that the atoms of species A are anticorrelated with themselves and with the atoms of species B, as both functions vanish along the diagonal. 
Note that at the same time atoms of species B are not strongly correlated. 
This is also captured by ansatz~(\ref{wavefunc0}), where strong correlations are induced by zeros whenever A-A or A-B atoms overlap (see panels (f) to (j)). 
All densities and pair correlations computed with this ansatz qualitatively resemble the exact correlation functions just before demixing. 
However, the ansatz fails to describe the ground state of the system once the system has phase separated.


Let us  note that the TBDF for the A species shown in Fig.~\ref{fig4}(c) corresponding to the crossover for $N_\mathrm{A}=N_\mathrm{B}=2$
look similar to those obtained for  $N_\mathrm{A}=4$ and a very heavy
atom in component B (discussed in~\cite{Pflanzer:2009,Pflanzer:2010})
or a large number of atoms in B (discussed in~\cite{Garcia-march:12}).
Those cases belong to the phase-separated limit, in which B formed
a material barrier. Therefore, the two atoms of A stay at each side of
B. Very differently in this case, there are only two atoms of A, and
they can be localized in either side of B.

\begin{figure}
\begin{center}
\hspace{-0.5cm}\includegraphics[scale=0.4]{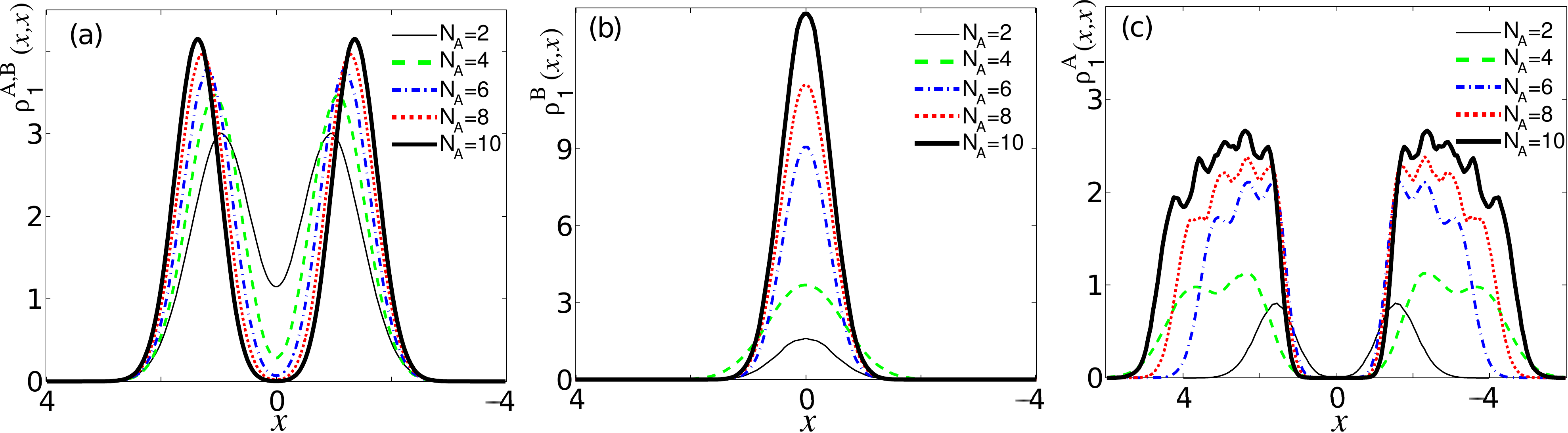} \\
\caption{(Color online)  {\it Densities with  $N_\mathrm{A}=N_\mathrm{B}=2,4,6,8,10$ atoms.} a) densities plotted in the composite fermionization limit, showing that the two peaks appear farther appart as $N$ is increased. b) densities for B in the phase-separated  limit. The atoms tend to localize more and more in the center as $N$ is increased. c)  densities for A in the phase-separated  limit. The atoms of A are in the edges of B, forming two TG gases with $N/2$ atoms. }  \label{Fig:DensitiesNBig}
\end{center}
\end{figure}

For $N_\mathrm{A}=N_\mathrm{B}>2$  the results discussed above remain qualitatively valid. We show in Fig.~\ref{Fig:DensitiesNBig}(a) the densities for the composite fermionization limit when $N_\mathrm{A}=N_\mathrm{B}=2,4,6,8,10$ calculated with DMC. In this situation, the OBDMs are equal for both species. The two peaks present in the density tend to spatially separate as $N$ is increased, as a consequence of the large repulsion between both species, which increases with the number of atoms.   In Fig.~\ref{Fig:DensitiesNBig}(b) and (c) we show the densities for B and A, respectively, in the phase-separated  limit. As $N$ is increased, the atoms of B have a greater tendency to localize in the center of the trap.  The numerically calculated density for A shows that this component is localized at each side of B, forming two TG gases with $N/2$ atoms in each side.

The difference in the energy between BEC-TG and TG-TG regimes is further increased in balanced systems of a larger size, $N_\mathrm{A}=N_\mathrm{B}\gg 1$. Indeed, according to Eq.~(\ref{E:BECBEC}), the energy in the BEC-TG scales linearly with the number of particles $N$, which is a typical behavior of weakly interacting bosons. Instead, in TG-TG limit according to Eq.~(\ref{E:TGBEC}) the dependence on $N$ is quadratic. The resembles the behavior of the energy of fermionic particles and is a manifestation of Girardeau mapping. Comparing the results for $N_\mathrm{A}=N_\mathrm{B}=2$ with $N_\mathrm{A}=N_\mathrm{B}=10$ we already observe how the difference in the energy between limits increases, see Fig.~\ref{Fig:Energies}.

\begin{figure}
\begin{center}
\includegraphics[scale=.8]{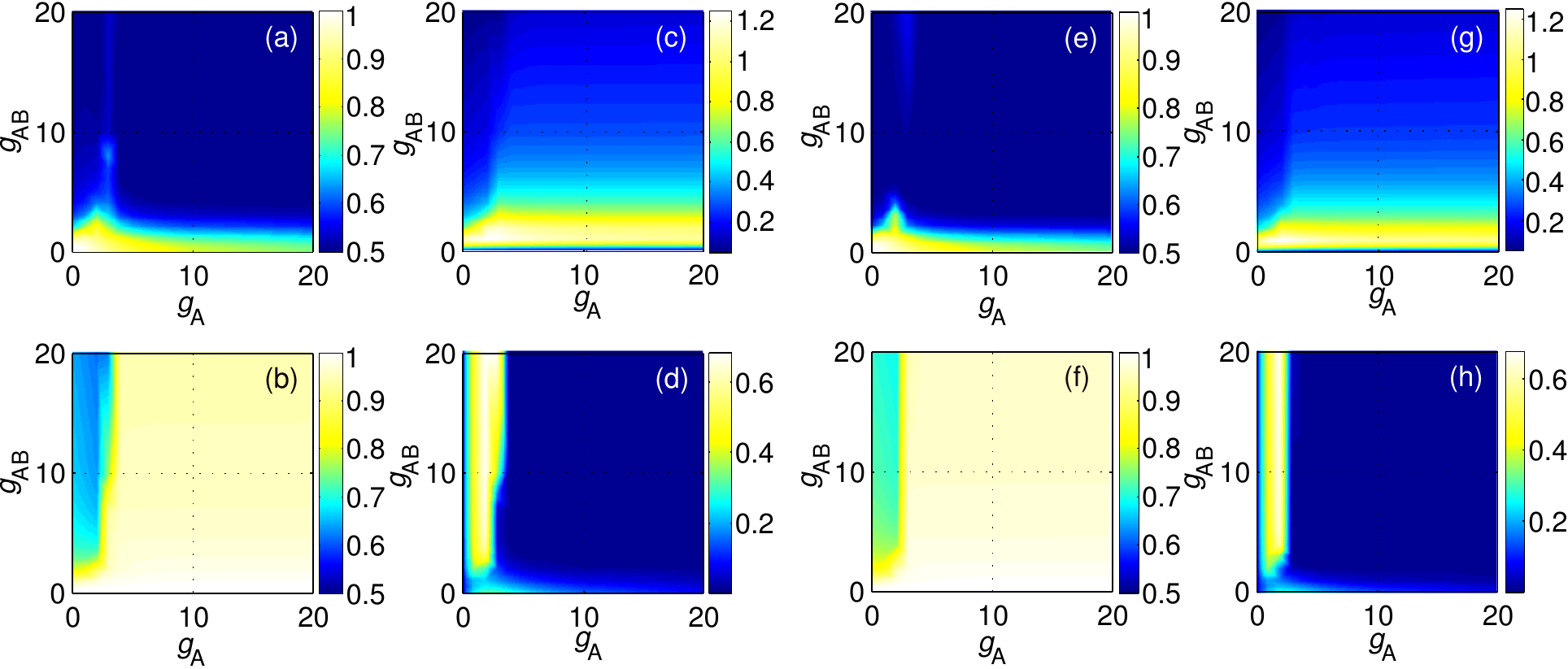}
\end{center}
\caption{(Color online) {\it  Largest occupations $\lambda_0^\mathrm{A}$  and $\lambda_0^\mathrm{B}$
of a natural orbital and average interaction energies as a function
of $g_\mathrm{A}$ and $g_{\mathrm{AB}}$, when $N_\mathrm{A}=2$,  $g_\mathrm{B}=0$. } (a)
and (b) show  $\lambda_0^\mathrm{A}$  and $\lambda_0^\mathrm{B}$  when  $N_\mathrm{B}=3$.
(c) and (d) represent $\langle U_{\mathrm{A}}\rangle\!$  and  $\langle U_{\mathrm{AB}}\rangle\!$,
respectively, for the same case. (e) to (h) represent the same when $N_\mathrm{B}=4$.
The region in which B is not condensed is reduced as  $N_\mathrm{B}$ is increased,
keeping $N_\mathrm{A}$ constant.  \label{fig7}}
\end{figure}

\section{Effect of a larger population in the weakly interacting species}
\label{Sec:weaklylarger}

In the imbalanced case, $N_\mathrm{B}>N_\mathrm{A}$, the wavefunction~(\ref{wavefunc0}) can be equally used as an ansatz for the exact ground state of the systems.
The four limits discussed above equally exist. 
Nevertheless, the weakly interacting species has now a greater tendency to localize in the center of the trap and condense, which modifies the boundaries between the different regimes associated to these limits.
In Fig.~\ref{fig7}(a)-(b) and (e)-(f) we report the largest eigenvalue of the OBDM for species A and B to quantify the coherence, covering the whole range of coupling constants, when $N_\mathrm{A}=2$ and $N_\mathrm{B}=3,4$, respectively.
As $N_\mathrm{B}$ is increased we observe that the region in which B is not condensed is reduced (the light blue area in Figs.~\ref{fig7} (b) and (f)). 
Moreover, the minimum value of $\lambda_0^\mathrm{B}$,
which occurs in this non-condensed area,  grows with $N_\mathrm{B}$ for fixed $N_\mathrm{A}$.    
Notice also that the area in which $\lambda_0^\mathrm{A}$ approaches the largest possible value $\lambda_0=1$, i.e. close to the $g_\mathrm{A}$ axis, is reduced as $N_\mathrm{B}$ is increased.

\begin{figure*}
\includegraphics[scale=0.43]{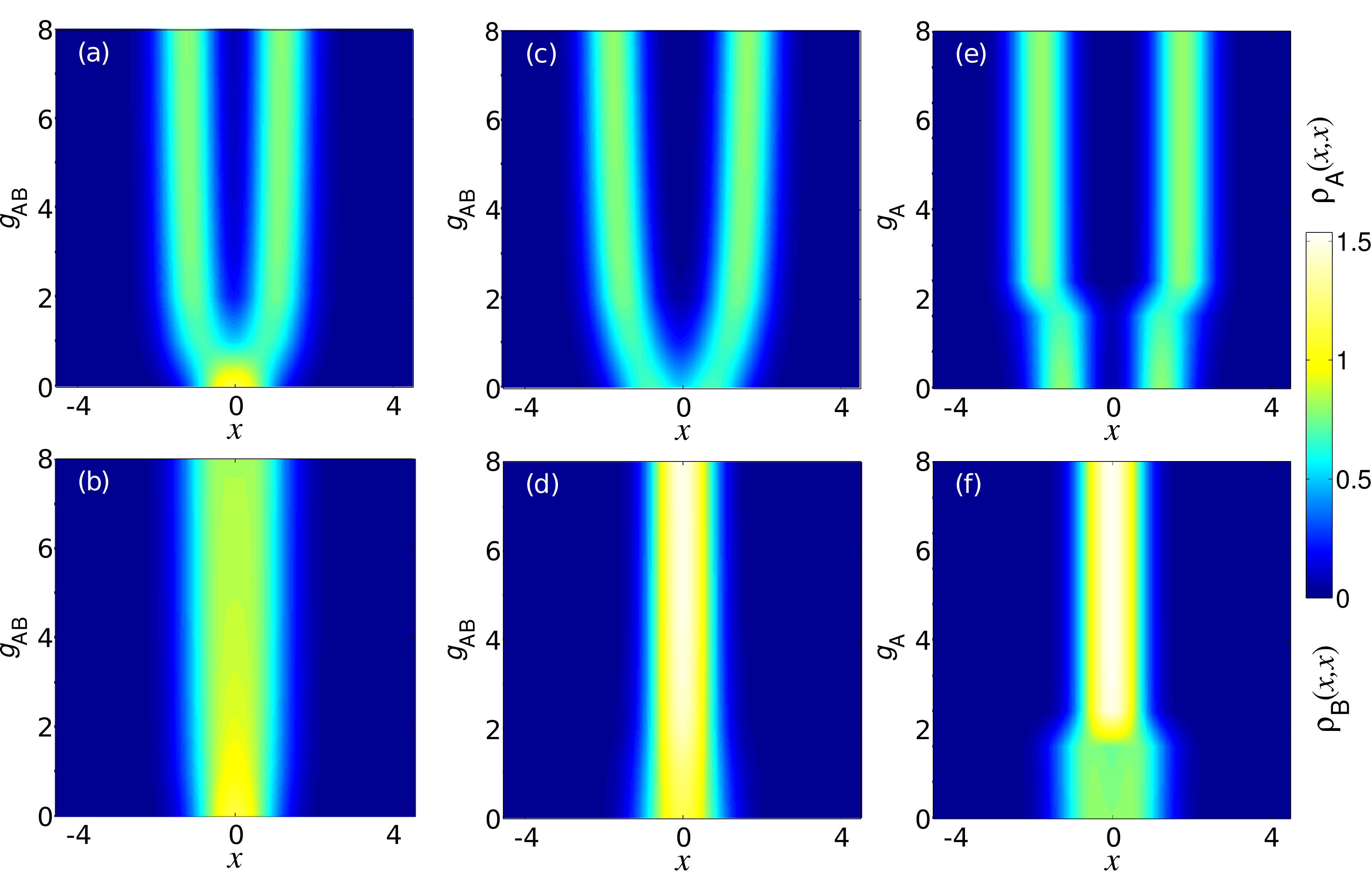}
\caption{(Color online) {\it Densities for both species between  the four
different regimes.}  Upper(lower) row  is the density for A (B)  species,
when $N_\mathrm{A}=2$ and $N_\mathrm{B}=4$.  Panel layout as in Fig.~\ref{Fig:Densities}.
The density for B in the composite fermionization limit is more similar to
the one obtained in the  phase separation limit.   \label{fig8}}
\end{figure*}

In Figs.~\ref{fig8} (a) and (b) we show the density profiles for A and B along the transition between the BEC-BEC limit and composite fermionization,
for $N_\mathrm{A}=2$ and $N_\mathrm{B}=4$. 
The atoms of species B are now more concentrated in the center than when both populations were equal, even though species B is still not fully condensed. 
The two peaks in species A appear at a smaller value of $g_\mathrm{AB}$, and are more spatially separated than in the case  $N_\mathrm{B}=N_\mathrm{A}$.  
We note that in the composite fermionization limit, the density of species A in the center for the balanced case is finite, while in the imbalanced case it vanishes (compare Figs.~\ref{Fig:Densities} (a) and Figs.~\ref{fig8} (a)). 
The density profiles along the transition between the TG-BEC  and the phase-separated gas are presented in Figs.~\ref{fig8} (c) and (d). 
Comparing with the balanced case plotted in Figs.~\ref{Fig:Densities} (c) and (d) we notice that, in the phase separation limit,  the two peaks in the density profile of A are now more separated and the squeezing in the density of B is smaller. 
The average interaction energy $\langle U_{\mathrm{AB}}\rangle\!$ (Fig.~\ref{fig7} (c) and (g)) tends to zero when phase separation occurs.
Figs.~\ref{fig8} (e) and (f) report the density along the transition between composite fermionization and phase separation. 
We observe that  the position of the two peaks in the density profile of A in the phase-separated and the composite fermionization limit is closer than in the balanced case (compare with Figs.~\ref{Fig:Densities} (e) and (f)).
Also, the crossover occurs now at a smaller value of $g_\mathrm{A}$. 
The average interaction energy $\langle U_{\mathrm{A}}\rangle\!$ (Figs.~\ref{fig7} (d) and (h)) decreases abruptly to zero after the crossover.
We conclude that for larger imbalances,  $N_\mathrm{B}\gg N_\mathrm{A}$, the composite fermionization region is highly suppressed, and therefore the surviving limits are those associated to BEC-BEC, TG-BEC and the phase-separated mixtures.

If the macroscopic limit is reached in such a way that the number of atoms in one of the species is fixed, the minority species plays role of an impurity which perturbs the majority species. The relative contribution of the minority species to the energy becomes smaller and polaronic description might be applicable. 

\label{Sec:conc}

Current experimental advances in ultracold atomic physics allow one to scrutinize the onset and evolution of correlations in few-atom bosonic fluids. 
Small samples can be trapped, and their interactions can be largely tuned, thus providing a fantastic ground to understand how quantum many-body correlations build in small samples. 
Binary mixtures are specially appealing as they provide the first step towards understanding the effect of environments on quantum systems in a controlled way. 
To advance in that direction, we study the effect of embedding a quantum fluid (component A) within a second quantum fluid (B) with tunable intra- and inter-species interactions at zero temperature. 
We fix the coupling constant of B-B interactions to that of ideal bosons, $g_\mathrm{B} = 0$, and vary A-A and A-B interactions in a wide range, 
$0\leq g_\mathrm{A}<\infty$, $0\leq g_\mathrm{AB}<\infty$. 
This permits us to explore the phase diagram for a variety of regimes. 
The energy, one- and two-body correlation functions, density profiles and von Neumann entropy are calculated exactly using diagonalization method. 
For larger system sizes, the results are complemented with the energy and density profiles obtained by diffusion Monte Carlo method. 

We have described the transition between the following four limits: a) BEC-BEC limit, where both components interact weakly and thus remain condensed, b) BEC-TG limit, where the two components interact weakly among each other and A has strong intra-species interaction, c) composite fermionization limit, where the interaction between both species is large, inducing strong correlations within both species, and d) a phase separation limit, where both the intra-species interaction in A and the inter-species interactions are large.  We show that the transition between the different limits involves sophisticated changes in the one- and two-body correlations. 
The energetic properties change in a smooth way, with the energy and its derivatives remaining continuous, which implies a transition of a crossover type rather than a true phase transition. At the same time, the entanglement between the two components has a much sharper dependence on the interactions.
This is demonstrated by reporting the von Neumann entropy, which manifests a sharp peak along the transition between composite fermionization and phase separation.
The evolution of the density profiles of A and B components is studied in detail both for the balanced and the imbalanced case.
The effect of a large number of particles on the energy and the density profiles is discussed. 
We analyze the coherence properties by expanding the one-body density matrix in natural orbitals and obtaining the occupation numbers.
We demonstrate that full condensation (largest occupation number equal to one) for A species is reached only in the BEC-BEC regime, while the weakly interacting
B species also remains fully condensed in the TG-BEC regime, and the
condensation is almost complete in the phase separation regime.
We argue that the described picture of the transition between four mentioned regimes remain valid also in a macroscopically large balanced mixtures, $N_\mathrm{A}=N_\mathrm{B}\to\infty$. 
Contrarily, when the macroscopic limit is reached by increasing the number of atoms of the weakly-interacting species, $N_\mathrm{B}\to\infty$, the composite fermionization limit is suppressed. 
Therefore the phase diagram in this highly imbalanced case resembles the one expected within a mean-field approach. 
The studied effects are relevant to ongoing and future experiments with small two-component systems.

\section{Acknowledgments}
  This project was supported by Science Foundation Ireland under Project
  No.~10/IN.1/I2979.  We acknowledge also partial financial support from the
  DGI (Spain) Grant No. FIS2011-25275, FIS2011-24154 and  the Generalitat de Catalunya Grant No. 2009SGR-1003. GEA and BJD are supported by the Ram\'on y Cajal program, MEC (Spain).

\end{document}